
\magnification=1200
\baselineskip=8truemm

\def\section{\vskip 2mm \centerline}
\def\r{\hangindent=1pc  \noindent}
\def\re{\hangindent=1pc  \noindent}
\def\cen{\centerline}
\def\v{\vskip 1mm}
\def\endpage{\vfil\break}
\def\noi{\noindent}
\def\kms{km s$^{-1}$}
\def\deg{$^\circ$}

\def\pa{PASJ}

\def\so{Sofue, Y.}

\cen{\bf DARK FILAMENTS IN THE GALAXY NGC 253}
\cen{\bf -- A BOILING GALACTIC DISK --}
\v\v

Yoshiaki  SOFUE$^1$, Ken-ichi WAKAMATSU$^2$, and David F. MALIN$^3$

{\it 1. Institute of Astronomy, University of Tokyo, Mitaka, Tokyo 181}

{\it ~~~~E-mail: sofue@mtk.ioa.s.u-tokyo.ac.jp}

{\it 2. Physics Department, Gifu University, Gifu 501-11, Japan}

{\it 3. Anglo-Australian Observatory, Epping NSW 2121, Australia}

\v\v\v
\cen{\bf Abstract}\v

We study the morphology of dark lanes and filaments in the  dust-rich galaxy
NGC 253 using an unsharp-masked $B$-band optical photograph. Dust features
are classified as `arcs', which have heights and scale radius of about 100
to 300 pc, connecting two or more dark clouds, and `loops' and `bubbles',
which are developed forms of arcs, expanding into the disk-halo interface.
These have diameters of a few hundred pc to $\sim 1$ kpc.
Among the bubbles, we notice a peculiar round-shaped bubble above the nucleus,
which could be a large-diameter ($\sim 300$ pc) supernova remnant exploded in
the halo over the nucleus. We also find
`vertical dust streamers',  which comprise bunches of narrow filaments with a
thickness of a few tens of pc and are almost perpendicular to the galactic
plane, extending coherently for 1 to 2 kpc toward the halo. Finally, we note
`short vertical dust filaments' (or spicules)
are found in the central region.
We interpret these features as due to
three-dimensional  structures of gas extending from the disk into the halo.
We propose a `boiling disk' model where the filamentary  features are
produced by star-forming activity in the disk as well as the influence of
magnetic fluxes.
We discuss the implication of the model for the chemical evolution of the
ISM in a galaxy disk.

\v\v

Subject Headings :Galaxies; individual (NGC 253) -- Galaxies: Halo -- ISM: dust
-- ISM: Magnetic fields

\section{1. INTRODUCTION}

Vertical dark lanes emerging from galactic disks are often recognized in
optical photographs of tilted spiral galaxies, and provide information about
the disk-halo interface and the magnetism (Sofue 1987; Sawa and Fujimoto 1987).
In addition to vertical lanes, numerous filaments are found in the
forms of loops (or bubbles) and arcs, which suggests that the galactic
disk is ``boiling and steaming'' (Sofue et al. 1991).
These features, particularly coherent bunches of straight vertical filaments,
are considered to be  deeply coupled with the magneto-hydro-dynamical (MHD)
activity in galactic disks and halos (Sofue and Fujimoto 1987; Sofue 1991).

In order to investigate such activity, we have undertaken a systematic
survey of dark filaments in the nearby Sculptor group galaxy NGC 253.  In
addition to its proximity [2.5 Mpc (Pence 1980)] this galaxy is
inclined at about 78 degrees to the line of sight and so is ideally
oriented for this kind of study. (Parameters for NGC 253 are listed
in Table 1.)
In addition, it is unusually dusty and
the dust can be seen in silhouette over most of the inclined disk.  Here,
we present the result of the survey and describe a morphological
classification  of the identified features. We further discuss implications
for the MHD activity of the various filamentary phenomena seen in the
disk-halo interface, and propose a ``boiling-disk" model.

\cen{- Table 1 -} \v

\section{2. DARK FILAMENTS IN NGC 253}

The optical photograph was taken at the F/3.3 prime focus of the
Anglo-Australian 3.9-n m telescope using the triplet corrector on
a IIIa-J plate combined with a GG385 filter.
The plate had been hypersensitised in nitrogen and hydrogen before
exposure, which was 80 minutes long and made (in a nitrogen atmosphere) on
1979 October 20-21 under excellent seeing conditions. In order to enhance
the small scale structures such as filaments, the plate had been copied
with the aid of an unsharp mask, which emphasizes the fine detail we are
interested in.
Fig. 1a, b and c show the photographs that resulted for the central region,
SW half and  NE half of the galaxy.
The center of the galaxy (nucleus position) is marked by the cross
in Fig. 1a.
On this photograph we notice numerous chaotic features both in emission
and the  dust lane absorptions in silhouette. We also present a composite
color photograph of NGC 253 in Fig. 2 which has been obtained from three
different black and white photographs made in the B, V, and R bands using
the same telescope.
The scales and positions can be identified by using stars 1 to 10, whose
positions are given in Table 2.

\cen{Table 2}

\cen{Fig. 1a, b, c; Fig. 2}

\section{\it 2.1. Spiral arms in emission and absorption}\v

Emission regions trace the bright spiral arms,  which comprise HII regions
and OB associations, seen as red  and blue knots (or clumps),
respectively in Fig. 2.  Using the unsharp-masked IIIa-J plate copy as well
as by referring to the color copy (Fig. 2), we made a sketch of extended
emission regions and HII regions. Fig. 3 shows the sketch we obtained
in this way.

In the same manner, we prepared a sketch of dark lanes, which  we show in
Fig. 4. Dark lanes are associated with the emission features,  tracing the
inner edges of the bright arms. The dark lanes are best seen on the unsharp
masked derivative of the IIIa-J plate, with the emission features clearly
tracing the inner edges of the bright arms. Although these features
trace the grand-designed spirals on a galactic scale, they are generally
clumpy and bumpy, or flocculent, and consist of  numerous clouds of
various sizes.

\cen{Fig. 3, 4}

\section{\it 2.2. Dark Filaments}\v

Numerous dust features are found emerging from the dark clouds,  apparently
``above'' the dark lanes and clouds along the spiral arms; Fig. 5 shows a
sketch of dark filaments which can be identified on the photograph. Their
appearance strongly suggests that the dusty filaments are structures that
are vertical with respect to the galactic plane, emerging towards the
disk-halo interface  and  halo. From their morphology, we have classified
the dusty filaments into `arcs', `loops' (or bubbles), vertical
`streamers', and `short vertical filaments' (or spicules).
These classifications are illustrated in Fig. 6.

\v\cen{Fig. 5, 6} \v

\noindent{2.2.1. Arcs}

Arcs are filaments with an arch-like structure, bridging two or more dark
clouds in the disk. They extend between 100 pc to 1 kpc in length, and
about 100 to 300 pc in height. Typical arcs are sketched in Fig. 7. They
are usually dome-like, but sometimes show a helmet-like shape, which is
suggestive of a relation to magneto-hydro-dynamical phenomenon such
as the Parker-type magnetic inflation.
The arcs are more often observed in the region of dense spiral arms
at galactocentric distances of 2-5kpc.

\cen{Fig. 7}

\noindent{2.2.2. Loops and Bubbles}

 Loops are more developed forms of arcs, and their radii are 200 pc to 1 kpc.
Some loops appear to be due to projected spherical bubbles, although their
three-dimensional structures are not definite from the photograph.
Fig. 8 reproduces a typical large-scale loop (bubble) found in
the north-eastern region, which appears to be expanding into the halo
with a diameter of about 0.7 kpc.
Besides the dust bubbles and loops, we have noticed an almost spherical
hole of dust in the central region, which we describe in detail in
section 2.3.

\cen{Fig. 8}

\noindent{2.2.3. Vertical Dust Streamer (VDS)}

These are open dust filaments running almost perpendicularly to the
galactic plane.
They are as thin as a few tens of pc or less (sometimes not
resolved with the  present 1 arcsec resolution), and lengths of 200 pc to 1
kpc.
Numerous vertical dust streamers are found  in the central region.
The largest VDS are often observed to extend 1 to 2 kpc from the disk,
and the largest streamers are found outside the nuclear region, especially
in the north-eastern region.
We show the largest streamers in Fig. 9, which is located 6$'$.4 (4.6 kpc)
toward the NE from the nucleus.
The coherent alignment of these vertical streamers suggests that the
filaments are somehow related to vertical magnetic tubes, along which
the low-temperature dust is accelerated without being dissociated.

\cen{Fig. 9}

\noindent{2.2.4.  Short Vertical Filaments}

There exist numerous vertical filaments with smaller scales something
similar to a brush,  very reminiscent of solar spicules.
Their shapes are similar to vertical streamers,
but scales are much smaller. They are as thin as ten pc or less, not
resolved by the present observations.
 Fig. 10 shows the central region where
short vertical filaments and numerous long vertical streamers are observed.

\cen{Fig. 10}

\v\cen{\it 2.3.  The Spherical  Bubble above the Nucleus}\v


In the central region, about 20$''$ ``above" the nucleus, we found an
almost perfect spherical structure, which appears as a void (hole) of dusty
filaments.
Fig. 11 shows this `central bubble', and its center position is given
in table 1.
The round edge of the hole is outlined by some enhanced dusty filaments.
Despite of its huge size (diameter of $\sim300$ pc), it shows no deformation.
A bright spiral arm can be traced across the bubble without any
interaction, which indicates that the arm is behind this bubble.
The velocity field as observed in the  CO  molecular line shows
no particular distortion and hence no interaction with the bubble
(Scoville et al 1985).
These facts suggest that the bubble is not in touch with the galactic
disk, but  is located in the halo.

\cen{ - Fig. 11 - }

Although the true position of this bubble along the line of sight is
ambiguous, the apparent proximity to the nucleus and its location on
the minor axis suggest that the bubble is just over the nucleus.
If this is the case, the height from the disk is estimated to be about
200 pc.
No counterpart to this object is found in the H$\alpha$ image
(Waller et al 1988).
No radio map with a sufficient resolution and coverage has been published,
so that we cannot investigate its radio properties.
An X-ray image at 0.2--4 keV observed with the Einstein Observatory
shows a feature extending  toward the SE along the minor axis in a good
positional coincidence with the bubble (Fabbiano, Trinchieri 1984).
This bubble object may be a huge spherical ejection from the nucleus.
In this context, we mention that the nucleus of NGC 253 is associated with
a `superwind' (Heckman et al 1991). It seems likely that the
bubble is connected with such a high-velocity outflow, although the
spherical shape appears difficult to be explained.
Alternatively, from its round and undisturbed appearance, it could be a
supernova remnant which has exploded in the halo.
However, there  still remains a question why the round shape can
be retained in spite of the interaction with the superwind.

\section{\bf 3. DISCUSSION}

\section{\it 3.1. `Boiling-and-Steaming' Disk}

The chaotic dusty features as illustrated in Fig. 3 lead us to propose
that the gaseous disk of NGC 253 is `boiling', and a part of the gas is
`steaming' into the halo.
Indeed, streaming motion has been observed along dusty jets in the
galaxy NGC 1808 (Phillips 1993).
However, understanding the formation mechanism
of the out-of-plane dusty filaments observed in NGC 253 remains a
challenging problem. The energy source of such turbulence may be the
injection of kinetic and thermal  energies from supernova explosions  as
well as by stellar winds. However, acceleration of dust grains up to the
heights of a few kpc, while their coherence is preserved, appears difficult
to be explained by a direct supply of energy from explosive events without
magnetic fields. On the other hand, the morphology of the loops and
arcs strongly suggests that they are somehow related to
magneto-hydro-dynamical phenomenon such as
the expansion of magnetic fluxes. The highly-aligned, coherent morphology
of the vertical dust streamers also suggests the presence of magnetic lines of
force running perpendicularly to the galactic plane. In the following, we
suggest possible mechanisms which might produce such coherent bunches of
dusty filaments, and discuss the energetics.

\section{\it 3.2. Energetics}

The average strength of magnetic fields in the disk of NGC 253 has been
estimated to be $13\pm 4$ $\mu$G (e.g., Sofue et al 1986). The field
strength in the halo will be weaker, and we may suppose that the field
strength in the halo-disk interface at height 100 to 1000 pc would be of
the order of one to a few $\mu$G. The magnetic energy involved in a typical
arc-shaped magnetic flux of strength 1 $\mu$G  with a radius 300 pc and
width 100 pc is, therefore, estimated to be $\sim 10^{50}$ erg, and $\sim
10^{51}$ erg for a field of 3 $\mu$G. The gravitational energy required to
lift a gaseous mass of the same volume with a density of $10^{-2} $ H
cm$^{-3}$ to a height of $h_z\sim 300$ pc is estimated to be
$E_{\rm grav} \sim m_{\rm gas} k_z h_z^2 \sim 10^{50}$ erg,
about the same as the magnetic energy,
where  $k_z$ is a constant related to the gravitational acceleration
$g_z(z)$  in the $z$ direction as $g_z(z) \simeq k_z z$,
and $h_z$ is the scale thickness of the stellar disk.
These amounts of energy are comparable to kinetic energy supplied
by a single supernova and/or by a stellar wind from a cluster of OB stars.
Hence, it will be energetically possible to produce the chaotic
three-dimensional dust features in NGC 253 by successive explosions of SN
in the disk as well as from stellar winds from massive stars.

\section{\it 3.3. Magnetic Flux Inflation}

It is possible that dust grains coexist with partially ionized gas  which
is entrained in a magnetic field. If the magnetic inflation
occurs in a time scale ($t_{\rm m}$) short enough  compared to the infall
time of the gas from the halo ($t_{\rm g}$),  dust grains can be carried to
the height by the inflating magnetic fluxes.
The time scales are given by
$$ t_{\rm m} \sim z/V_{\rm A} \sim \sqrt{4 \pi \rho} z/B
$$
$$\sim 10^7 {z \over {300 ~{\rm pc}}}
\left({\rho \over {0.01 m_{\rm H} {\rm cm}^{-3}}}\right)^{1/2}
\left({B \over {10^{-6}{\rm G}}}\right)^{-1}
 ~{\rm yr}, \eqno (1) $$
where $V_{\rm A}$ is the Alfven velocity,
$z$ is the height from the galactic plane,
$\rho$ is the density of gas in the magnetic tube,
and $B$ is the field strength.
If we take
$\rho \sim 10^{-2} m_{\rm H} {\rm cm}^{-3}$
and $B \sim 10^{-6}$ G, we have
$t_{\rm m} \sim 10^7$ yr for $z\sim 300$ pc.

The gas infall time is given by
$$ t_{\rm g} \sim 1/\sqrt{k_z} \sim h_z/v_z
 \sim 10^7 {h_z \over {300 ~{\rm pc}}}
\left({v_z \over {30 ~{\rm km~s^{-1}} }}\right)^{-1} ~{\rm yr}, \eqno(2) $$
where  $v_z$ is the velocity dispersion of disk stars in the $z$ direction.
For a normal galactic disk we may take
$h_z \sim 300$ pc
and $v_z \sim 30$ \kms,
then we have $t_{\rm g} \sim 10^7$ yr.
Hence, if the magnetic strength is greater than 1 $\mu$G and/or
the gas density within magnetic tubes is less than some
$\sim 0.01 m_{\rm H}{\rm cm}^{-3}$, raising the dust grains to
the observed height is possible.

\section{\it 3.4. Acceleration by Radiation-Pressure along Magnetic Tubes}

The radiation pressure due to starlight from the disk and bulge,
particularly from starburst regions, would act to
drive (accelerate) dust grains along vertical magnetic tubes, producing
vertical dusty structures (Ferrini et al. 1991).
The time scale of a grain to be accelerated to a height $z$ by the radiation
pressure is roughly estimated by
$$ t_{\rm rad} \sim \sqrt{a \rho_{\rm d} z c / I} $$
$$ \sim 2\times 10^7 \left({a \over {1 \mu {\rm m}}}\right)^{1/2}
\left({\rho \over {1 {\rm g~cm^{-3}}}}\right)^{1/2}
\left({z \over {300 ~{\rm pc}}}\right)^{1/2}
 \left({I \over {10^{44} {\rm erg~s^{-1}} /(\pi {10~{\rm kpc}^2})}}
\right)^{-1/2}  ~ {\rm yr}, \eqno(3)$$
where $a$ and $\rho_{\rm d}$ are the size (radius) and density of the grain,
respectively, $c$ is the light velocity, and $I$ is the intensity
of star light which can be approximated by $I \sim L/(\pi R^2)$
with $L$ and $R$ being the luminosity and effective
radius of the galaxy disk, respectively.
If we take $a \sim 1 \mu{\rm m}$,
$\rho_{\rm d} \sim 1$ g cm$^{-3}$, $z \sim 300$ pc, $L \sim 10^{44}$
ergs s$^{-1}$,
and $R \sim 10$ kpc, then we have $t_{\rm rad} \sim 2\times 10^7$ yr.
This is in the same order as the infall time, and, therefore,
the acceleration of dust grains is quite possible.

In order for the magnetic lines of force to act as guide lines ( or tubes)
for dust grains blown by stellar radiation pressure, the strength of the
magnetic field must be strong enough,  or  $t_{\rm mag} \ll t_{\rm rad}$. The
radiation pressure as well the magnetic field strength in the central few kpc
region of the galaxy would be much stronger than the values assumed above,
and, therefore, the acceleration due to the radiation pressure would be
more effective in the central region than in the outer region. In fact, the
vertical streamers and short filaments  are more often observed in the
central region.

\v\cen{\it 3.5. Implications for the Chemical Evolution of the ISM}\v

We have obtained the morphological classification of dark lanes and
filaments in the galaxy NGC 253. The features are naturally understood if
they are out-of-plane structures, emerging from the galactic plane toward
the halo, and appear to establish a disk-halo interface. We have suggested
some possible mechanism for the formation of the vertical structures and
given arguments about the energetics involved. However, individual features
are too complicated to be explained by such simple mechanisms, and we have
to wait for a detailed theoretical modeling, particularly by taking into
account MHD activities in the disk-halo interface. It is also necessary to
obtain further information about kinematics and quantitative estimates of
the masses and magnetic fields involved in the features.

The outflow of gas from the galactic disk has implications
for the circulation of interstellar gas.
Particularly, such an energetic ejection as the large-scale vertical dust
streamers
will play a significant role in lifting the `metal-polluted'
(or metal-enriched) interstellar gas into the halo.
The gas lifted to the halo will then fall back to the disk at different
locations, causing a  mixing of the ISM.
If the gas is blown off far away from the central region by the radiation
pressure due to strong starlight from the nuclear region,
it will fall on the outer disk.
This would then result in a galactic scale mixing (circulation) of dusty
gas from the inner to outer disk.
This will inevitably result in a galactic-scale mixing of heavy elements,
and therefore, a smearing out the metallicity gradient with respect to
galacto-centric distance.
The increase in metallicity enhances the
formation of dust and molecular clouds, so
the large-scale circulation of the metallicity results in the increase
of the star-formation rate in the outer disk:
The boiling-and-steaming disk phenomenon would, therefore, affect the
chemical evolution of a disk galaxy.

\section{References} \v

\r Fabbiano, G., Trinchieri, G. 1984, ApJ, 286,  491.

\re Ferrini, et al 1991, in  The Interstellar Disk-Halo Connection in
 Galaxies, IAU Symp 144, ed. J. B. G. M. Bloemen (Kluwer, Dordrecht), p. 397.

\r Heckman, T. M.,  Armus, L.,  Miley, G. K. 1991, ApJS 74 833

\re Pence, W.D. 1980, ApJ, 239, 54

\re Philips, A. 1993, AJ, 105, 486

\re Sawa, T., Fujimoto, M.  1987, in  Magnetic Fields and
Extragalactic Objects, ed. E.Asseo and D. Gre{\'e}sillon (Edition de Phys.),
pp. 165-169.

\re Scoville, N. Z., Soifer, B. T., Neugebauer, G., Young, J., Yerka, J.
1985, ApJ, 299, 129.

\re \so 1991, in {\it The Interstellar Disk-Halo Connection in Galaxies, IAU
Symp 144}, ed. J.B.G.M. Bloemen (Kluwer Academic Publishers, Dordrecht),
p. 169.

\re Sofue, Y. 1987, PASJ, 39, 547

\re Sofue, Y.,  Fujimoto, M. 1987, \pa,  39, 843

\re Sofue, Y., Fujimoto, M., Wielebinski, R. 1986, ARAA, 24, 459.

\re \so, Y., Wakamatsu, K.,  Malin, D. F. 1991, in {\it The Interstellar
Disk-Halo Connection in Galaxies, IAU Symp 144}, ed. J. B. G. M. Bloemen (
Kluwer Academic Publishers, Dordrecht), p. 309.

\r Waller, W. H., Kleinmann, S. G.,  Ricker, G. R. 1988, ApJ, 95, 1057.

\endpage

\settabs 2 \columns
\centerline{Table 1. Parameters for NGC 253}
\v

\hrule \vskip 1mm \hrule \v

\+ Position of the nucleus$^\dagger$ \dotfill & RA = 0h 45m 05.9s \cr
\+ 	& Dec = $-25$\deg 33$'$ 40$''.1$ \cr
\+ Distance \dotfill  & 2.5 Mpc (Pence 1980) \cr
\+ Node P.A. \dotfill  & 47\deg \cr
\+ Inclination \dotfill & 78\deg \cr

\v \hrule

\v
\noi $\dagger$ NED (NASA Extragalactic  Database, IPAC) 1994.
\vskip 20mm

\settabs 10 \columns

Table 2.  Equatorial Coordinates (1950.0) of the reference stars around
NGC 253
\vskip 1mm
\hrule
\vskip 1mm
\+  Star && RA(h ~m ~s)&&  ME($''$)& Dec($^{\circ} ~~ ' ~~''$) && ME($''$)&
$l(^{\circ})$ & $b(^{\circ})$ \cr
\vskip 1mm
\hrule \vskip 0.5mm \hrule
\vskip 1mm

\+ 1 \dotfill &&00 44 28 77.70 &&0.08&-25 39 10.58 &&0.15 & 92.69 &-87.98  \cr

\+ 2 \dotfill &&00 44 54 59.80 &&0.07&-25 39 45.41 &&0.15  &95.00 &-88.04   \cr

\+ 3 SAO166575 \dotfill && 00 44 52 89.4 &&&     -25 39 13.48 \cr

\+ 4 \dotfill && 00 44 48.403 &&0.07&-25 34 21.96&& 0.31&  95.58& -87.95  \cr

\+ 5  \dotfill &&00 44 48.327 &&0.08&-25 33 52.40 &&0.00 & 95.68& -87.94  \cr

\+ 6 \dotfill  &&  00 45 16.100 &&0.31&-25 33 59.57 && 0.07 & 98.30& -87.99 \cr

\+ 7 \dotfill   && 00 45 36.184 &&0.15&-25 32 25.33 &&0.00& 100.58 &-87.99  \cr

\+ 8 \dotfill   &&00 45 43.467 &&0.08&-25 29 58.89 &&0.08 &101.72& -87.96  \cr

\+ 9 \dotfill  &&00 45 33.263 &&0.07&-25 26 56.98 &&0.15 &101.23& -87.90  \cr

\+ 10 \dotfill   &&00 45 45.395 &&0.08&-25 27 27.45 &&0.08 &102.32 &-87.93  \cr

\+ Central Bubble \dotfill  &&00 45 07.007 &&0.23&-25 33 52.68 &&0.15  &97.45
&-87.97  \cr

\vskip 1mm
\hrule
\vskip 1mm
\noi Star 3 was used for the reference of the positions.
The distance between stars 3 and 6 is 7$'$.400 = 5.38 kpc for
a distance of 2.5 Mpc of the galaxy; the distance between stars 2 and 5
is 6$'$.05 = 4.40 kpc.


\endpage

\noi{Figure Captions}\v\v\v

\r  Fig. 1. Unsharp-masked photograph of the galaxy NGC 253.
The photograph was taken at the F/3.3 prime focus of the 3.9m Anglo-Australian
Telescope using the triplet corrector on a IIIa-J plate combined with a
GG385 filter. (a) The central region of the galaxy; (b) the SW half;
and (c) the NE half.
Positions of stars 1 to 10 referred to the SAO star (star 3) are given
in table 2. The cross in Fig. 1(a) marks the position of the nucleus.

\v\v

\r Fig. 2. Natural-color photograph of NGC 253, as obtained from three
different color photographs using the same telescope as for Fig. 1.

\v\v

\r  Fig. 3. A sketch of emission regions made from Fig. 1. The
emission regions trace spiral structures.
\v\v

\r Fig. 4. A sketch of dark lanes, which align predominantly along the spiral
arms. \v\v

\r Fig. 5. A sketch of dark filaments, which comprises dark arcs, loops and/or
bubbles, vertical streamers, and short dust filaments. \v\v

\r  Fig. 6. Classification of the major filamentary features into arcs, loops
and/or bubbles, long vertical streamers, and short vertical filaments.
\v\v

\r Fig. 7: Arcs.
\v\v

\r  Fig. 8: Loops and bubbles.
\v\v

\r  Fig. 9: The largest-scale vertical dust streamers in the NE.
\v\v

\r  Fig. 10: Short vertical filaments (spicules) in the central region.
The cross marks the position of the  nucleus.

\v\v

\r Fig. 11: The spherical bubble (hole) `above' the nucleus (marked by the
cross). The diameter of the bubble is approximately 300 pc.
The bar indicates $30''$=366 pc.

\bye